\documentclass[10pt]{article}

\usepackage{twocolumn}
\usepackage{epsf}
\usepackage{jeep}
\mysection{section}{\normalsize\bf}{\thesection.~}   
\mysection{subsection}{\normalsize\it}{\thesubsection.~}

\newcommand{\text}[1]{\mathrm{#1}}

\newcommand{\dst}{|\Delta \text{S}|\text{=2}}
\newcommand{\kkm}{${\text{ K}^0\!-\!\overline{\text{K}^0}}\,$-mixing\/}
\newcommand{\kkmd}{${\text{K}_{\text{L}}\!-\!\text{K}_{\text{S}}}\,$-mass
difference\/}
\newcommand{\gev}{\, \text{GeV}}
\newcommand{\laMSb}{\Lambda_{\overline{\text{MS}}}^{\text{NLO}}}

\begin{document}
\title{{\normalsize TUM-T31-92/95 \hfill hep-ph/9510383 } \\[3mm]
PHENOMENOLOGY OF $\varepsilon_K$ BEYOND LEADING
LOGARITHMS\thanks{Talk presented at the EPS-HEP 95 conference,
Brussels, Jul.~27-Aug.~2 1995.
}}
\author{\textsc{Ulrich Nierste}\thanks{supported by BMBW under
contract no.~06-TM-743}\\[2mm] \textit{Physik-Department,
Techn.~Univ.~M\"unchen, D-85747 Garching, Germany} \\[4mm]
\parbox{15truecm}{\small
I present the QCD short distance coefficient $\eta_3$ of the $\Delta
S=2$ hamiltonian in the \emph{next-to-leading order} (NLO) of
renormalization group improved perturbation theory.  Since now all
QCD-factors $\eta_1$, $\eta_2$ and $\eta_3$ are known with NLO
accuracy, a much higher precision in the analysis of $\varepsilon_K$
can be achieved.  The CKM phase $\delta$, $|V_{td}|$
and the mass difference $\Delta
m_{B_s}$ in the $\text{B}_s^0-\overline{\text{B}}_s^0$-system
are predicted from the measured values of $\varepsilon_K$ and
$\Delta m_{B_d}$. Finally I briefly look at the \kkmd.
This work has been done in collaboration with Stefan
Herrlich.}  }
\date{}
\maketitle
\maketitle
\pagestyle{empty}

\section{The $\dst$-hamiltonian in the NLO}
\thispagestyle{empty}
The low-energy hamiltonian inducing \kkm\/ reads:
\begin{eqnarray}
H^{|\Delta S|=2} =
               \frac{ G_{F}^2 }{ 16 \pi^2 } M_W^2  \left[
                  \lambda_c^2 \eta_1^{\star}
                   \frac{m_c^{\star\,2}}{M_W^2}  \! + \!
                  \lambda_t^2 \eta_2^{\star}
                  S(\frac{m_t^{\star\,2}}{M_W^2} ) \nonumber \right. \\
\quad \quad  \! + \!  \left.
                2 \lambda_c \lambda_t \eta_3^{\star}
                  S(\frac{m_c^{\star\,2}}{M_W^2} ,
                    \frac{m_t^{\star\,2}}{M_W^2}  )
                   \right]
 b(\mu) Q_{S2}(\mu) + \text{h.c.} \; \; \label{s2}
\end{eqnarray}
Here $G_{F}$ is the Fermi constant, $M_W$ is the W boson mass,
$\lambda_j=V_{jd} V_{js}^{*}$ comprises the CKM-factors and $Q_{S2} $
is the local $\dst$ four-quark operator
\begin{eqnarray}
Q_{S2} &=&
( \overline{s}_j \gamma_\mu (1-\gamma_5) d_j)
(\overline{s}_k \gamma^\mu (1-\gamma_5) d_k)  \label{ollintro}
\end{eqnarray}
with $j$ and $k$ being colour indices.
$m_q^{\star}=m_q (m_q)  $, $q=c,t$, are running quark masses in the
$\overline{\text{MS}}$ scheme.
The Inami-Lim functions $S(x),S(x,y)$  describe the
$|\Delta \text{S}|\text{=2} $-transition amplitude in the absence of strong
interaction.

The short distance QCD corrections are comprised in the coefficients
$\eta_1$, $\eta_2$ and $\eta_3$ with a common factor $b(\mu)$ split
off.  They are functions of the charm and top quark masses and of the
QCD scale parameter $\Lambda_{\text{QCD}}$.  Further they depend on
the \emph{definition} of the quark masses used in the Inami-Lim
functions: In (\ref{s2}) the $\eta_i$'s are defined with respect to
$\overline{\text{MS}}$ masses $m_q^{\star}$ and are therefore marked with
a star.

With actual values of the input data the results of the old
\emph{leading log approximation} read
\begin{eqnarray}
\eta_1^{\star \,\text{LO} } \approx 0.80 \, ,
\quad \quad
\eta_2^{\star \,\text{LO} }  \approx   0.62 \, ,
\quad \quad
\eta_3^{\star \,\text{LO} }  \approx   0.36 \, .
\label{old}
\end{eqnarray}
Now the NLO values read:
\begin{eqnarray}
\eta_1^{\star} = 1.32 \!\!
\begin{array}{l}
\scriptstyle +0.21 \\[-.7mm]
\scriptstyle -0.23                 
\end{array}, \, \, \,
\eta_2^{\star}  =  0.57 \! \!
\begin{array}{l}
\scriptstyle +0.00 \\[-.7mm]
\scriptstyle -0.01
\end{array}, \, \, \,
\eta_3^{\star}  =   0.47 \! \!
\begin{array}{l}
\scriptstyle +0.03 \\[-.7mm]
\scriptstyle -0.04                 
\end{array} , \label{etas}
\end{eqnarray}
where $m_c^{\star}=1.3 \gev$ and $\laMSb=0.310 \gev$ has been used.
$\eta_2$ and $\eta_3$ are almost independent of the input parameters.

The NLO calculation of $\eta_2$ has been performed in \cite{bjw} and
the NLO result for $\eta_1$ can be found in \cite{hn1}. $\eta_3$
in (\ref{etas}) is new. Details of the calculation are presented in
\cite{hn86}. A  phenomenological analysis using the NLO $\eta_i$'s
has been published in \cite{hn3}.

(\ref{old}) and (\ref{etas}) clearly show the sizeable numerical
effect of the NLO correction. Further there are conceptual reasons
for going to the NLO:
\begin{enumerate}
\renewcommand{\labelenumi}{\roman{enumi})}
\item The fundamental QCD scale parameter
      $\Lambda_{\overline{\text{MS}}}$ cannot be used in LO calculations.
\item The quark mass dependence of the $\eta_i$'s is not
      accurately reproduced by the LO expressions.  Especially the
      $m_t$-dependent terms in
      $\eta_3 \cdot S(m_c^{\star\,2}/M_W^2, m_t^{\star\,2}/M_W^2 )$
      belong to the NLO.
\item Likewise the proper definition of the quark masses is
      a NLO issue: One must go to the NLO to learn how to use
      $m_t$  measured at FERMILAB in low energy expressions such as
      (\ref{s2}). In NLO the $\overline{\text{MS}}$ mass $m_t^{\star}$ is
      smaller than the \emph{pole mass}  $m_t^{\text{pole}}$ by
      8$\gev$.
      It has been discussed at this conference to which definition
      of $m_t$ the quoted CDF and D0 results for $m_t$ refer.
      Presumably the measured  quantity is $m_t^{\text{pole}}$.
\item In the NLO the large LO error bars caused by renormalization scale
      dependences are reduced.
\end{enumerate}

\section{Phenomenology of $\varepsilon_K$}
Let us first recall our present knowledge about the CKM matrix $V_{CKM}$:
The precise  measurements of $|V_{ud}|$ and $|V_{us}|$ also constrain
$|V_{cd}|$, $|V_{cs}|$ and $|V_{tb}|$ via the unitarity of $V_{CKM}$.
$|V_{cb}|$ is expected to lie in the range
\begin{eqnarray}
0.037 \leq |V_{cb}| \leq 0.043 \label{vcb}
\end{eqnarray}
After fixing $|V_{cb}|$ unitarity likewise  pins down $|V_{ts}|$.
We will further need
\begin{eqnarray}
0.06 \leq \frac{|V_{ub}|}{|V_{cb}|} \leq 0.10 \label{vub} .
\end{eqnarray}
Yet even for fixed $|V_{cb}|$ and $|V_{ub}/V_{cb}|$ the magnitude of
the remaining CKM element $|V_{td}|$ is a sensitive function of
the phase $\delta$. Hence $\Delta m_{B_d}$ and $|\varepsilon_K|$
provide complementary information on $|V_{td}|$:
The former determines this element directly  and the latter
indirectly through $\delta $.

Apart from $|V_{cb}|$ and $|V_{ub}/V_{cb}|$ two other key parameters
are involved: The actual value  $m_t^{\text{pole}}=(180 \pm 12) \gev$
for the top quark pole mass corresponds to
\begin{eqnarray}
160 \gev \leq m_t^{\star} \leq 184 \gev \label{mt} .
\end{eqnarray}
Finally the hadronic matrix element of $Q_{S2}$ in (\ref{ollintro})
is parametrized as
\begin{eqnarray}
\langle \overline{\text{K}^0} | Q_{S2} (\mu) |  \text{K}^0 \rangle &=&
\frac{8}{3} f_K^2 m_K^2 B_K / b( \mu ) .
\nonumber
\end{eqnarray}
We take
\begin{eqnarray}
0.65 \leq B_K \leq 0.85         ,  \label{bkrange}
\end{eqnarray}
which reflects the $1/N_c$ result as well as the ballpark of the
lattice determinations.

The uncertainties in the input parameters other than $|V_{cb}|$,
$|V_{ub}|/|V_{cb}|$, $m_t$ and $B_K$ do not significantly affect the
analysis (see \cite{hn3}).  Using the LO $\eta_i$'s, however,
imposes an error onto $\varepsilon_K$, which is  roughly of the
same order as the one due to (\ref{vcb}-\ref{bkrange}). The NLO shift
in $\eta_3$ affects $\varepsilon_K$ as much as pushing $B_K$ from
0.75 to 0.82. This uncertainty has been neglected in most
phenomenological analyses.

\begin{figure}[htb]
\centerline{
\epsfxsize=8.2cm
\epsffile{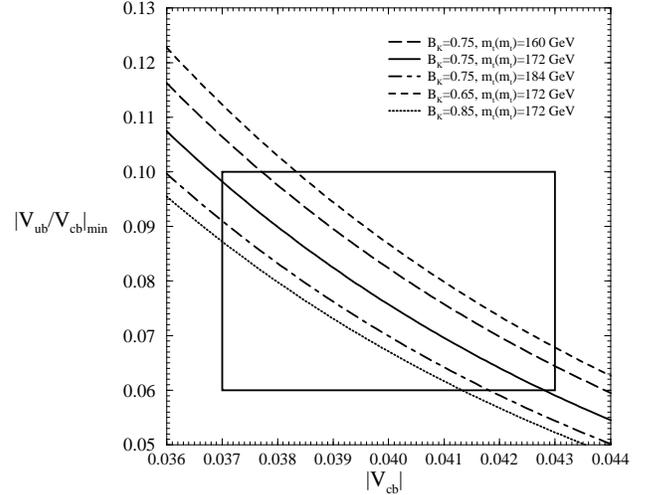}
}
\caption{For each pair $(m_t^{\star},B_K)$ the measured value for
$\varepsilon_K$ defines a curve. The points below the curve are
excluded.  The rectangle limits the allowed range for $|V_{cb}|$ and
$|V_{ub}/V_{cb}|$ obtained from tree-level b-decays according to
(\ref{vcb}) and (\ref{vub}).
}\label{vmin}
\end{figure}

After fixing three of the key parameters the measured value for
$\varepsilon_K$ yields a lower bound at the fourth one. This feature
has been used in the pre-top era to constrain $m_t$. Yet today one
should focus on the CKM elements instead. The allowed range for
$(|V_{cb}|,|V_{ub}/V_{cb}|)$ is shown in fig.~\ref{vmin}.

Next $\delta$ is obtained from a simultaneous analysis of
$\varepsilon_K$ and $\Delta m_{B_d}$:
For $|V_{ub}|/|V_{cb}|=0.08, |V_{cb}|=0.04, m_t^\star=172 \gev,B_K=0.75 $
one finds the two solutions:
\begin{eqnarray}
\delta^{\text{low}} = 85^{\circ}, \,\, &&
\delta^{\text{high}} = 121^{\circ}   .
\end{eqnarray}
They correspond to
\begin{eqnarray}
\left|V_{td}\right|^{\text{low}}  =
9.1 \cdot 10^{-3}, \, \, &&
\left|V_{td}\right|^{\text{high}} =
10.7 \cdot 10^{-3} .
\end{eqnarray}
Accounting for  the errors (\ref{vcb}-\ref{bkrange}) leads to
\begin{eqnarray}
47^{\circ} \leq \delta \leq 135^{\circ}, &&
7.3  \leq   |V_{td}| \cdot 10^{3} \leq 11.9 ,
\label{range}
\end{eqnarray}
where the upper bounds stem from
$\Delta m_{B_d}=(0.470\pm 0.0025) \,ps^{-1}$. Here
$F_{B_d} \sqrt{B_{B_d}} = ( 195 \pm 45 ) \, \text{MeV}$ has been used.

The ratio $\Delta m_{B_d}/\Delta m_{B_s}$ is theoretically much better
understood than the mass differences separately.  We can use the
result  (\ref{range}) for $|V_{td}|$ to predict $\Delta m_{B_s}$:
\begin{eqnarray}
6.2 \, ps^{-1} \leq   \Delta m_{B_s} \leq 26 \, ps^{-1} .
\end{eqnarray}
More details can be found in \cite{hn3}, where slightly different
values for $m_t$ and $\Delta m_{B_d}$ have been used.

\section{The ${\mathbf{K}_{\mathbf{L}}\!-\!\mathbf{K}_{\mathbf{S}}}\,$-mass
difference}
In the 1980s it was generally believed that long distance
interactions make up more than half of the observed   \kkmd.  Today
the picture has changed due to the NLO enhancement of $\eta_1$ and
$\eta_3$ and the larger value  for $\Lambda_{\overline{\text{MS}}}$,
which pushes $\eta_1$ further up. One finds
\begin{eqnarray}
\frac{\left(\Delta m_K\right)_{\text{SD}}}
        {\left(\Delta m_K\right)_{\text{exp}}}
&=& 0.7 \pm 0.2 \label{sd}
\end{eqnarray}
exhibiting a short distance dominance. This is in agreement with naive
power counting: One expects the long distance part to be suppressed
with a factor of $(\Lambda_{had}/m_c)^2$ compared to (\ref{sd}). Here
$\Lambda_{had}$ is a hadronic scale.

\end{document}